# Direct observation of Landau level resonance and mass generation in Dirac semimetal $Cd_3As_2$ thin films

Xiang Yuan[1,2,†], Peihong Cheng[1,2,†], Longqiang Zhang[3,4], Cheng Zhang[1,2], Junyong Wang[5], Yanwen Liu[1,2], Qingqing Sun[6], Peng Zhou[6], David Wei Zhang[6], Zhigao Hu[5], Xiangang Wan[3,4], Hugen Yan[1,2,*], Zhiqiang Li[7,*], Faxian Xiu[1,2,*]

[1] State Key Laboratory of Surface Physics and Department of Physics, Fudan University, Shanghai 200433, China

[2] Collaborative Innovation Center of Advanced Microstructures, Fudan University, Shanghai 200433, China

[3] National Laboratory of Solid State Microstructures, School of Physics, Nanjing University, Nanjing 210093, China

[4] Collaborative Innovation Center of Advanced Microstructures, Nanjing University, Nanjing 210093, China

[5] Key Laboratory of Polar Materials and Devices, Ministry of Education, Department of Electronic Engineering, East China Normal University, Shanghai 200241, China

[6] State Key Laboratory of ASIC and System, Department of Microelectronics, Fudan University, Shanghai 200433, China

[7] College of Physical Science and Technology, Sichuan University, Chengdu, Sichuan 610064, China

[†] These authors contributed equally to this work.

[*] Correspondence and requests for materials should be addressed to F. X. (E-mail: faxian@fudan.edu.cn), Z. L. (E-mail: zhiqiangli@scu.edu.cn) and H. Y. (E-mail: hgyan@fudan.edu.cn).

**Abstract**

**Three-dimensional topological Dirac semimetals have hitherto stimulated unprecedented research interests as a new class of quantum materials.[1-5] Breaking certain types of symmetries has been proposed[1,2] to enable the manipulation of Dirac fermions; and that was soon realized by external modulations[6-8] such as magnetic fields. However, an intrinsic manipulation of Dirac states, which is more efficient and desirable, remains a significant challenge. Here, we report a systematic study of quasi-particle dynamics and band evolution in $Cd_3As_2$ thin films with controlled Chromium (Cr) doping by both magneto-infrared spectroscopy and electrical transport. For the first time, we observe $\sqrt{B}$ relation of inter-Landau-level resonance in undoped $Cd_3As_2$ Dirac semimetal, an important signature of ultra-relativistic Dirac state inaccessible in previous optical experiments. A crossover from quantum to quasi-classical behavior makes it possible to directly probe the mass of Dirac fermions. Importantly, Cr doping allows for a Dirac mass acquisition and topological phase transition enabling a desired dynamic control of Dirac fermions. Corroborating with the density-functional theory calculations, we show that the mass generation is essentially driven by the explicit $C_4$ rotation symmetry breaking and the resultant Dirac gap engineering through Cr substitution for Cd atoms. The manipulation of the system symmetry and Dirac mass in $Cd_3As_2$ thin films provides a tuning knob to explore the exotic states stemming from the parent phase of Dirac semimetals.**

Three-dimensional (3D) topological Dirac semimetals constitute a novel topological phase with linear energy dispersions along all three momentum directions.[3, 4, 9] In Dirac semimetals, the conduction and valence bands contact with each other only at discrete points, which consist of degenerated Weyl nodes with opposite chiralities.[1] Recent theoretical calculation[5] suggests that $Cd_3As_2$, a generally perceived Kane-type semiconductor[10], is intriguingly a Dirac semimetal with a linear electronic dispersion in all three dimensions. This proposal was soon confirmed by the experimental observation of Dirac nodes through photoemission spectroscopy.[4, 9, 11] Landau quantization of the 3D Dirac quasi-particles has been studied by scanning tunneling microscopy[8] and transport experiments.[6, 12, 13] Particularly, the ultra-high mobility and giant magnetoresistance (MR) in $Cd_3As_2$[12] have fueled substantive research interests towards a wide spectrum of applications.

Theoretically, Dirac semimetals have been predicted to be the parent phase of many exotic states such as Weyl semimetals, topological insulators, massive Dirac states and topological superconductors, by breaking certain symmetries.[1, 2, 5, 14-23] Extrinsic modulations in $Cd_3As_2$ involving high pressure[24] and magnetic field[6-8] have been proved to be capable of tuning the band structure or the carrier dynamics. A plenty of exciting phenomena such as chiral anomaly[25] and unconventional superconductivity[26, 27] were observed in $Cd_3As_2$. Nevertheless, these modulations may require certain extreme conditions, like point stress or high magnetic field. An intrinsic control of the massless Dirac state, as an efficient and practical way, is highly desired. Theoretical calculations and numerical simulations have also proposed interesting physics based on element doping such as rich spin textures[28], Kondo effect[29], spin correlation[30] and topological state with room-temperature ferromagnetic order[31]. However, these predictions are yet to be experimentally investigated or verified.

In this study, molecular-beam epitaxy (MBE) technique is employed to produce high-quality $Cd_3As_2$ thin films. MBE has been widely conceived as a powerful tool for precise control of doping concentration in the study of topological insulators.[32-34] Here, we conduct a systematic investigation of quasi-particle dynamics and band evolution in $Cd_3As_2$ thin films with different Cr concentrations by magneto-infrared spectroscopy and electrical transport. Without Cr doping, the inter-Landau-level resonance exhibits a $\sqrt{B}$-dependence, a benchmark of ultra-relativistic Dirac fermions in $Cd_3As_2$. Reducing the magnetic field leads to a crossover from quantum to quasi-classical behavior that is characterized by a linear $B$ dependence of the cyclotron resonance[35]. Combined with transport measurements, an accurate extraction of quasi-particle mass has been achieved. Remarkably, by introducing Cr dopants into the $Cd_3As_2$ lattice, a novel phase transition is triggered along with mass generation in the absence of a long-range ferromagnetic order. Thus, the $Cd_3As_2$ system can be tuned from a massless to a massive Dirac state prior to the complete elimination of the topological phase. Further theoretical calculations unveil that the mass acquisition is driven by explicit $C_4$ rotation symmetry breaking which results in a gap generation. Our work establishes a feasible way to manipulate the 3D Dirac fermions through the controllable element doping.

The presence of magnetic field will lead to electron orbit quantization and the formation of highly-degenerated Landau levels. In conventional materials with parabolic dispersion, the Landau levels are uniformly spaced on the energy scale[36] as described by $E_n = (n+1/2)\hbar eB/m^*$, where $n, \hbar, e, m^*$ is Landau index, reduced Plank constant, elementary charge and cyclotron mass, respectively. The frequency of cyclotron resonance is given by $\omega_c = eB/m^*$. In Dirac semimetals, the linear dispersion results in an unequally spaced Landau level spectrum[37-41] following $E_n = \text{sgn}(n)\sqrt{2v_F^2 eB\hbar n}$ as shown in Fig. 1a, where $v_F$ denotes the Fermi velocity. The existence of Dirac point will give an unconventional zero mode ($n = 0$) that doesn't shift with magnetic field. Optical transitions between Landau levels lead to the minima in the transmission spectrum; and in addition to the inter-band transitions which are marked by purple arrows in Fig. 1a, intra-band transitions (red arrows) are also likely to be observed as long as they follow the dipole selection rules $\Delta n = \pm 1$.[37, 42, 43]

In order to investigate the band dispersion, we performed transmission spectroscopy measurements of $Cd_3As_2$ thin films grown on semi-insulating GaAs substrates. The growth was monitored by *in-situ* reflection high energy electron diffraction (RHEED) under ultra-high vacuum. A typical RHEED image is shown in Fig. 1b inset where the streaky patterns can be clearly resolved, revealing a flat surface morphology and single crystallinity. The film thickness was specifically designed to be ~200 nm-thick for the moderate infrared transmittance, and also, to completely remove quantum size effect (refer to Supplementary section X). The X-ray diffraction (XRD) peaks can be indexed as series of {112} planes (Fig. 1b). The magneto-transmission spectra were recorded in the Faraday geometry at liquid helium temperature and normalized by the zero-field spectrum (Fig. 1c). Infrared light was focused on the sample through a parabolic cone and the transmitted light was subsequently detected by a bolometer. Fig. 1d exhibits the magneto-transmission ratio spectra at different magnetic fields. The curves are vertically shifted for clarity. The blocked area comes from GaAs reststrahlenband which has no transmission. Several transmission minima are clearly observed and they show a systematic shift to higher energy with increasing magnetic field. These minima correspond to the optical absorption maxima due to the inter-Landau-level transitions. At high fields, two transmission minima can be resolved while they gradually merge together with the decrease of magnetic field. Another set of transmission minima with much narrower width (position ~ 200 cm$^{-1}$ at 16 T), is determined to come from GaAs substrate itself (refer to Fig. S2).

Firstly we focus on the high magnetic field regime ($B$ > 10 T) where two transmission minima exist. The observation of two minima suggests the existence of Landau levels which are not uniformly distributed on the energy scale. The possibility of Zeeman splitting can be safely ruled out by performing high-field magneto-transport measurements (refer to Supplementary section XI). In order to quantitatively analyze the high-field behavior, the wavenumber of the transmission minima was extracted using Lorentz fitting (Fig. S3) and plotted versus $\sqrt{B}$ as shown in Fig. 2a. For each set of minima energy, they are linearly proportional to $\sqrt{B}$ and the intercept crosses zero point. This behavior is in stark contrast to the conventional quadratic dispersion systems, where Landau levels are uniformly spaced and the resonance energy is proportional to $B$.[36] It, however, can be well understood in the context of Dirac system, where linear energy dispersion and Dirac point result in the observed $\sqrt{B}$-dependence and zero intercept. As illustrated in Fig. 2c, electrons can only be excited from occupied states below Fermi energy to empty states above Fermi energy (marked by red arrows). Given spatially inhomogeneous Fermi energy $E_F$, other nearby transitions are also allowed (marked by blue arrows).[44] When the magnetic field is high and the transition index is fixed, the resonance energy should be proportional to $\sqrt{B}$ as well. The ratio of the two sets of minima energy in Fig. 2a is calculated to be 1.19, very close to $(\sqrt{3}-\sqrt{2}):(\sqrt{4}-\sqrt{3})$. Thus we can assign the resonance at higher energy (red rectangles) to be the transition from $L_2$ to $L_3$ ($L_n$ represents the $n^{th}$ Landau level), whereas the resonance at lower energy (blue rectangles) to be $L_3$ to $L_4$. The Fermi velocity $v_F$ is extracted as $1.6 \times 10^6$ m/s. All other possible resonances regardless of Fermi energy can be marked by grey lines in Fig. 2a.

The allowed resonance energy and the transition index experience a sudden change as the Landau level crosses the Fermi energy. The corresponding resonance can be theoretically modeled based on the Dirac-type Landau level distribution equation with independent parameters of Fermi velocity $v_F$ and cyclotron mass $m^*$ as attained from Fig. 2a and b, respectively. All the experimental data can be well fitted to the model, *i.e.*, the red and blue solid lines in Fig. 2a and b. When decreasing the magnetic field, several jumps can be resolved from the theoretical plot (the red and blue solid lines in Fig. 2b). After each jump, the Landau level index is increased by one and the resonance energy suddenly decreases to a smaller value, but they still follow the $\sqrt{B}$ relation for each fixed index. If further reducing the magnetic field, the Landau level will frequently cross the Fermi level especially when the Fermi level is far away from Dirac point. As shown in Fig. 2a, the platform-like resonance energy will evolve into a smooth curve below 5 T. The resonance energy between adjacent Landau levels follows $\hbar\omega_c = \sqrt{2v_F^2 B\hbar}(\sqrt{n+1}-\sqrt{n})$. With high Fermi level and low magnetic field, it can be approximated by $\hbar\omega_c \sim \sqrt{2v_F^2 eB\hbar}\frac{1}{2\sqrt{n}} = \frac{v_F^2 eB\hbar}{E_F} = \frac{e\hbar B}{m^*}$. Hence, under such circumstance, a classical cyclotron-resonance-like behavior can be observed even in the Dirac system. Using the magneto-optical data from the low fields (Fig. 2b), we can calculate $m^* = 0.02m_e$, where $m_e$ is the free electron mass. As shown in Fig. 2d, the frequent change of transition index will lead to a distinct quasi-linear-$B$ dependence of intra-band excitation, which perfectly matches our experimental results (Fig. 2b).

Dirac semimetals have been predicted to be the parent phase of many intriguing states, which were partially realized by recent transport experiments under high magnetic field[6, 45]. As an efficient way of tuning the band topology and realizing strong modulations of electronic states, Cr doping was performed by precisely co-evaporating Cd, As and Cr elemental sources in a MBE system. The doping concentration $x$ is defined by the atomic ratio through energy dispersive X-ray spectroscopy (EDX). All the films have the same thickness of ~200 nm. Fig. 3a presents the normalized magneto-infrared absorption spectra for a Cr-doped thin film with the doping concentration $x \sim$ 2.1 %. Only one set of transmission minima can be resolved. Some irregular features near ~ 570 cm$^{-1}$ remain unidentified. However, since their frequencies do not evolve with magnetic field, they are unlikely associated with the cyclotron resonance. The resonance energy increases with magnetic field and overlaps with the GaAs reststrahlenband at 12 T. The minima positions for the undoped and 2.1% Cr-doped Cd$_3$As$_2$ thin films are plotted in Fig. 3b. Both of them show a quasi-classical linear relation with $B$ in low fields. The slope of the Cr-doped sample is much smaller, corresponding to a larger cyclotron mass of $0.05\, m_e$; that is more than twice the value of the undoped sample. No clear resonance behavior is observed in samples with higher Cr-doping concentration, presumably due to a lower mobility, a further enhanced cyclotron mass and a smaller resonance energy beyond the experimental detection limit (Supplementary section VII).

We further carried out the magneto-transport measurements on the as-grown thin films. As illustrated in Fig. 4a, conventional multi-terminal Hall bar devices were used for transport experiments with a typical constant current of 0.5 μA. At liquid helium temperature, a positive MR is detected under perpendicular magnetic field (Fig. 4b). For the undoped sample, a comparatively large MR ratio and quasi-linear relation of the MR are observed as the general feature of undoped 3D Dirac semimetals.[12, 46] As increasing the doping concentration, the MR ratio is found to be strongly suppressed. Meanwhile, the quasi-linear MR gradually evolves to a parabolic curve. Clear

Shubnikov-de Haas (SdH) oscillations are witnessed to be superimposed on the MR background. The carrier cyclotron mass can be also extracted from the temperature-dependent SdH oscillations. As presented in Fig. 4c, the normalized oscillation amplitude $\Delta\rho/\rho(0)$ of the samples with different Cr concentrations shows a systematic trend in accordance with the doping percentage. The oscillation amplitude follows the formula $\Delta\rho(T) = \Delta\rho(0)\lambda(T)/\sinh(\lambda(T))$ with thermal factor $\lambda(T) = 2\pi^2 k_B T m^*/\hbar eB$, where $k_B$ is Boltzmann's constant. By performing the best fit, $m^*$ in different samples is found to increase from 0.02 $m_e$ (x ~ 0%) to 0.24 $m_e$ (x ~ 5.9%), showing a good agreement with the magneto-optical results (Fig. 4d). It is clear that by introducing a small amount of Cr dopants, $m^*$ dramatically increases by an order of magnitude, from which we may anticipate a significant modification of band structure induced by the Cr doping.

To probe the change of band dispersion, we also calculate the Berry phase $\phi_B$ based on the Landau fan diagram (refer to Fig. S8), which is a fundamental parameter in condensed matter systems. Landau quantization follows the Lifshitz-Onsager quantization rule, $A_F \frac{\hbar}{eB} = 2\pi(n+\gamma) = 2\pi(n+\frac{1}{2}-\frac{\phi_B}{2\pi})$, where $A_F$ is the cross-sectional area of Fermi surface related to the Landau index $n$. The intercept $\gamma$ of Landau fan diagram is quantitatively linked with Berry phase through the equation $\gamma \equiv 1/2 - \phi_B/2\pi$. Unconventional Dirac systems such as graphene[47, 48] and topological insulators[49, 50] share the same nontrivial $\pi$ Berry phase that is resulted from the persistent zero mode under magnetic field. Here, we plot the magnetic field of peaks (valleys) versus integer (half-integer) Landau index in the fan diagram (Fig. 4e). The intercept on *n*-axis shows a systematic development in accordance with the Cr concentration. The undoped $Cd_3As_2$ thin film displays a perfect zero offset ($\gamma=0$), equivalent to π Berry phase (Fig. 4f), which is a clear indication for the existence of Dirac node and non-trivial topological state. This is in agreement with $\sqrt{B}$ dependence of inter-Landau-level resonance in the magneto-optical experiments. It is worth noting that *m\** of the $Cd_3As_2$ is nearly doubled with 2.1% doping.

In order to understand the mass generation and tuning effect by Cr doping in $Cd_3As_2$, several possible mechanisms in our scenario are discussed as follows, involving (i) the change of Fermi area in the massless picture based on the discussion of quasi-linear *B* response in magneto-optical measurements; (ii) the variation of band curvature once the Cr-doping shifts the Fermi energy out of the linear E(k)-dispersion regime; and finally (iii) Dirac gap engineering in the massive Dirac fermion case. In order to clearly identify each contribution, we must revisit the Onsager equation. The Fermi area $A_F$ is linear to the *nB* term, thus it can be experimentally estimated by the slope of the fan diagram. As shown in Fig. 4e, in the light doping region ($x \leq 4.6\%$), all three curves exhibit nearly an identical slope. We can therefore conclude that the Fermi area or Fermi vector $k_F$ is not significantly changed and the mass acquisition from (i) and (ii) is negligible. For the scenario (iii), instead of Fermi area variation, the band dispersion itself is possibly tuned by the Cr doping. A direct expectation for the mass generation in linear dispersion system is that the Dirac system experiences a phase transition from the massless Dirac state to the massive Dirac state, where a quasi-particle gap is generated to separate the upper and lower branches of the Dirac cone. Dispersion in this scenario is given by $E = \sqrt{\hbar^2 v_F^2 k^2 + \Delta^2}$, where the $\Delta$ is the mass term. The Dirac gap has been considered as a prerequisite of many interesting phenomena such as quantum anomalous Hall and magneto-electric response.[51] And the gapped features have been observed in several massive Dirac systems such as topological (crystalline) insulators[15-20] and graphene[52-54] which are induced by breaking certain symmetries. The Fermi level in $Cd_3As_2$ always lies in the conduction band, impeding the direct observation of quasi-particle gap. As shown later, theoretical calculations were performed to further support the Dirac gap opening and existence of massive Dirac fermions induced by the Cr doping.

At low energy of massive Dirac fermion, the dispersion is no longer linear and the Berry's phase becomes path dependent in stark contrast to the massless case.[55] If we only account for the dispersion contribution to the cyclotron orbit, the offset $\gamma$ in massive Dirac states will be neither zero (massless Dirac fermion) nor 1/2 (normal massive fermion). However, a recent study[56] clearly pointed out that the offset $\gamma$ in the Onsager equation also includes the contribution from the pseudospin magnetization component. Taking the extra contribution to the phase offset into account will exactly cancel the original one induced by the pure band dispersion. Therefore, even though there is a phase transition from massless to massive states, the offset $\gamma$ is supposed to maintain zero due to the opposite contribution of the band dispersion and pseudospin magnetization. This phenomenon is protected by so-called particle-hole symmetry.[57] Experimentally, however, we found a distinctive $\gamma$ value of 0.3 in 2.1%-doped $Cd_3As_2$, which obviously deviates from zero (massless Dirac fermion). The co-existence of finite $\gamma$ and Dirac gap suggests a more complex band with particle-hole symmetry breaking.[55] As a result, the band dispersion consists of a deviation from the ideal massive Dirac fermion and the electron/hole pocket become asymmetric. Instead of simply incorporating a Dirac mass in the dispersion, the particle-hole asymmetric quadratic term should be added, written as $\Delta E(k) = \frac{\hbar^2}{2m}k^2$, which results in non-ideal Dirac states. Such non-ideal Dirac fermions have been observed in $Bi_2Te_2Se$.[55, 58] They can also contribute to the generation of cyclotron mass and in turn lead to a deviation of $\gamma$ from zero.[55] Therefore, a more realistic picture will be the coexistence of the aforementioned two mechanisms. Both the Dirac and quadratic terms contribute to the mass acquisition. The Hamiltonian is therefore formulated as $H = (\frac{\hbar^2 k^2}{2m} - \mu) + \begin{pmatrix} \Delta & \hbar v_F(k_y + ik_x) \\ \hbar v_F(k_y - ik_x) & -\Delta \end{pmatrix}$ and the dispersion relation changes to $E = \pm\sqrt{\hbar^2 v_F^2 k^2 + \Delta^2} + \frac{\hbar^2 k^2}{2m}$, where $\mu$ is Fermi energy. As discussed by the previous work[18], SdH oscillations are limited for the direct extraction of $\Delta$. For a more detailed and quantitative understanding of the Cr tuning effect, we need to consider the symmetry of the system and also to perform density functional theory as discussed in the following context.

In principle, both Dirac and quadratic masses are tunable by controlling the doping in the light doping regime (x ≤ 4.6%). Further increasing Cr dopants saturates the offset to 0.5 (Fig. 4e, green and blue lines) and directly produces a zero Berry phase (Fig. 4f). The heavy Cr doping exceeding 4.6% results in a non-topological phase and the elimination of Dirac point. Other effects of heavy doping involve the change of Fermi surface, which is beyond the current scope of this study since the phase transition has already been completed. As the consequence of the phase transition, the carrier mobility also drops quickly with increasing doping concentration. The linewidth in the magneto-optical experiments is significantly broadened and the magneto-transport measurements also quantitatively confirm this observation (for details, refer to Supplementary section VII). The remarkable change in the quasi-particle dynamics within 5.9% of Cr suggests that the system is highly sensitive to the Cr dopants.

Since the massive Dirac states are generally realized by breaking certain symmetries in massless Dirac fermions systems, such as breaking time reversal symmetry in topological insulators[16, 18, 19], mirror symmetry in topological crystalline insulators[15, 17] and sublattice symmetry in graphene[52-54], it is also natural to ask what the underlying mechanism is in the current 3D Dirac semimetal $Cd_3As_2$. Since Cr is a popular candidate in magnetic doping, Cr doping may also lead to the question whether there is a ferromagnetic order developed in the system and if the topological phase transition is induced by the magnetic doping owing to the time reversal symmetry breaking. Our results, however, elucidate that the system does not carry a long-range ferromagnetic order as the magneto-transport and superconducting quantum interference device (SQUID) experiments do not reveal any sign of hysteresis (refer to Supplementary section XI).

To further understand the physical mechanism and evaluate the contributions of different mass, we performed density-functional theory (DFT) calculations on both undoped and Cr-doped $Cd_3As_2$. The crystal structure of $Cd_3As_2$ is body-centered tetragonal and the Dirac point is protected by fourfold rotational symmetry.[5] We used the experimental lattice parameters (a=b=12.67Å; c=25.48Å) but relaxed the internal coordinates. Experimentally, we found that the Cd is substituted by Cr through EDX measurements (Supplementary Table S1). In pure $Cd_3As_2$, there are 6 distinct nonequivalent position types for Cd, and each type contains 8 different positions (denoted as a1-a8 in Supplementary Table S2) in a primitive unit cell (Fig. 5a). First, we calculate the electronic structure of the pure $Cd_3As_2$ (blue point in Fig. 5b), which agrees very well with the preceding reports[5, 25], showing a Dirac cone along $\Gamma - Z$ line. Then we substitute one Cr for one Cd atom in a primitive unit cell, corresponding to the concentration of $x \sim 2\%$. Configurations with the Cr atom in the same position type is equivalent, so it has only 6 nonequivalent configurations here. We calculate the total energy of doped $Cd_3As_2$ with the Cr atom in non-equivalent position a1-f1, respectively. As shown in Supplementary Table S2, the system with the Cr atom in position f1 is most stable. The relative energy of these configurations is quite large. One can expect that the Cr atom occupies position f1 (Fig 5a) instead of distributing randomly. Thus, the substitution of Cr atoms will automatically break the $C_4$ rotational symmetry around $k_z$ axis. Without the rotational symmetry protection, the Dirac nature and band topology are significantly altered.

Fig. 5c displays the electronic structure of Cr-doped Cd$_3$As$_2$ with one of the Cd atoms in unit cell substituted by a Cr atom (f1 position). We compare two band structures inside the red boxes in Fig. 5b and c. Due to the broken rotation symmetry, the original Dirac cone along $\Gamma - Z$ line is eliminated by the generation of a finite quasi-particle gap which directly supports the existence of massive fermions. On the other hand, the offset in the Landau fan diagram suggests the existence of particle-hole symmetry breaking. Therefore, combining the DFT calculations and the experimental results, the observed mass generation in Cd$_3$As$_2$ is contributed by both the Dirac mass and particle-hole asymmetry. We further carried out the calculations for the proposed Hamiltonian following the procedure from Ref.(57). Based on the experimental parameters $v_F, \gamma$ and the DFT-calculated gap of $2\Delta \sim 50\text{meV}$, we estimate the quadratic mass generation to be 0.008$m_e$ in 2% doped sample, which is much smaller than the overall cyclotron mass generation 0.03$m_e$. Based on the above discussions, we understand that the quadratic mass term was induced by the Cr doping which accounts for the finite offset, but the overall mass generation upon the doping is mainly contributed by the Dirac mass from the Dirac gap opening. In short, slight Cr doping is able to introduce Dirac mass, particle-hole symmetry breaking, and, to strongly modify the band dispersion. Adding more atoms into the system further lowers the crystal symmetry and enhances the influence from the Cr band. The calculated results generally agree with the experimentally-observed mass generation, as a direct consequence of the $C_4$ rotation symmetry breaking of the system.

In conclusion, we provide a direct spectroscopic determination of Dirac fermions in $Cd_3As_2$ thin films by magneto-optics, along with the observation of classical-quantum resonance crossover. Remarkably, the phase transition and Dirac mass acquisition can be achieved by the Cr doping. Our DFT calculations reveal that the Cr substitution leads to the $C_4$ rotation symmetry breaking which results in Dirac gap and band topology engineering. The controllable Dirac mass that we have achieved in the representative Dirac semimetal $Cd_3As_2$ opens up a feasible path towards the manipulation of exotic states stemming from the parent phase of Dirac semimetals.

**Method**

**Thin Film Growth.**

A series of $Cd_3As_2$ thin films were grown in a CREATEC MBE system with base pressure lower than $2 \times 10^{-10}$ mbar. GaAs substrates were degassed at 350 °C for 30 min to remove any molecule absorption prior to the growth. A very thin layer of CdTe below 5 nm was introduced as buffer and it has been proved not to affect the transport or optics measurements. The $Cd_3As_2$ thin film deposition was carried out by co-evaporating high-purity Cd (99.999%) and As (99.999%) from dual-filament and valve-cracker effusion cells, respectively. The beam flux ratio Cd/As was fixed around 3, and the growth process was *in-situ* monitored by RHEED. For Cr doping, another cell with Cr (99.999%) was used for co-evaporation and the doping concentration was precisely controlled by adjusting the cell temperature.

**Magneto-optical measurements.**

The far-infrared transmittance was measured in a Faraday configuration (perpendicular field) with a superconducting magnet. The sample was exposed to the infrared light through light pipes. Infrared light was focused with parabolic cone, detected by a bolometer and analyzed by a Fourier transform infrared spectrometer (FTIR). All the light tube, samples and bolometer were kept at liquid helium temperature in a cryostat. The light path was pumped under vacuum to avoid the absorption of water and other gases.

**Magneto-transport measurements.**

Magneto-transport measurements were carried out at different temperatures using a physical property measurement system with magnetic field up to 9 T. A standard Hall-bar geometry is made. Stanford Research 830 Lock-in amplifiers were used to measure the electrical signals.

**Density-functional theory calculations.**

The first-principles electronic structure calculations were performed within the generalized gradient approximation (GGA) of Perdew, Burke and Ernzerhof (PBE) as implemented in Vienna Ab initio Simulation Package (VASP), using the projector augmented wave (PAW) method. The energy cutoff was taken to be 350 eV. A 3×3×3 k-point grid was used for Brillouin zone integration. We used the experimental lattice parameters but relaxed the internal coordinates. The convergence criterion of structural relaxation was that the Hellman-Feynman forces on the ions were less than 0.006 eV/Å. No signal of magnetic order was found in experiment, thus we performed non-magnetic calculation. Spin-orbit coupling was also included in our calculations.

**Acknowledgements**

This work was supported by the National Young 1000 Talent Plan, the Program for Professor of Special Appointment (Eastern Scholar) at Shanghai Institutions of Higher Learning, and National Natural Science Foundation of China (61322407, 11474058 and 11525417). Part of the sample fabrication was performed at Fudan Nano-fabrication Laboratory. A portion of this work was performed at the National High Magnetic Field Laboratory, which is supported by National Science Foundation Cooperative Agreement No. DMR-1157490 and the State of Florida.

**FIGURES**

**Figure 1 | Landau-level fan chart, material characterizations and magneto-optical spectra of the undoped Cd$_3$As$_2$ thin film. a**, Unevenly spaced Landau levels in Dirac semimetal as a function of magnetic field. Arrows show the allowable optical transitions. **b**, Typical XRD spectra, showing {112} crystal planes and good crystallinity. The inset is an *in-situ* RHEED image of the Cd$_3$As$_2$ thin film. **c**, Schematic plot of the magneto-optical experiment setup. **d**, Normalized magneto-infrared spectra under different magnetic fields.

**Figure 2 | Crossover from quantum to classical resonance and Landau levels in the undoped Cd$_3$As$_2$ thin film. a,b**, Resonance wavenumber plotted against $\sqrt{B}$ and *B*, respectively. The red and blue curves represent theoretically modeling based on $v_F$ and *m\**. Grey lines show each set of adjacent Landau level transitions or the quasi-classical cyclotron resonance. **c,d**, The Landau levels and allowable transitions in the quantum regime and the quasi-classical regime, respectively.

**Figure 3 | Magneto-optical observation of mass generation in Cr-doped Cd$_3$As$_2$ thin films. a**, Normalized magneto-infrared spectra of 2.1% Cr-doped Cd$_3$As$_2$. **b**, Quasi-classical cyclotron resonance, indicative of the mass generation in the doped samples.

**Figure 4 | Magneto-transport observation of mass generation in Cr-doped Cd$_3$As$_2$ thin films. a**, A schematic drawing of magneto-transport setup. **b**, MR at 2.5 K for the undoped and Cr-doped Cd$_3$As$_2$ under perpendicular magnetic field. **c**, Normalized SdH oscillations amplitude plotted with temperature for different doping concentrations. **d**, Cyclotron mass generation upon Cr doping measured by both transport and optics. **e**, Landau fan diagram. **f**, The Berry phase evolution from π to zero with the increase of Cr.

**Figure 5 | Band structure calculations based on density-functional theory. a**, The crystal structure and atom position in x ~ 2% doped Cd$_3$As$_2$. **b,c**, Calculated band structures of the undoped and ~ 2% Cr-doped Cd$_3$As$_2$, respectively.

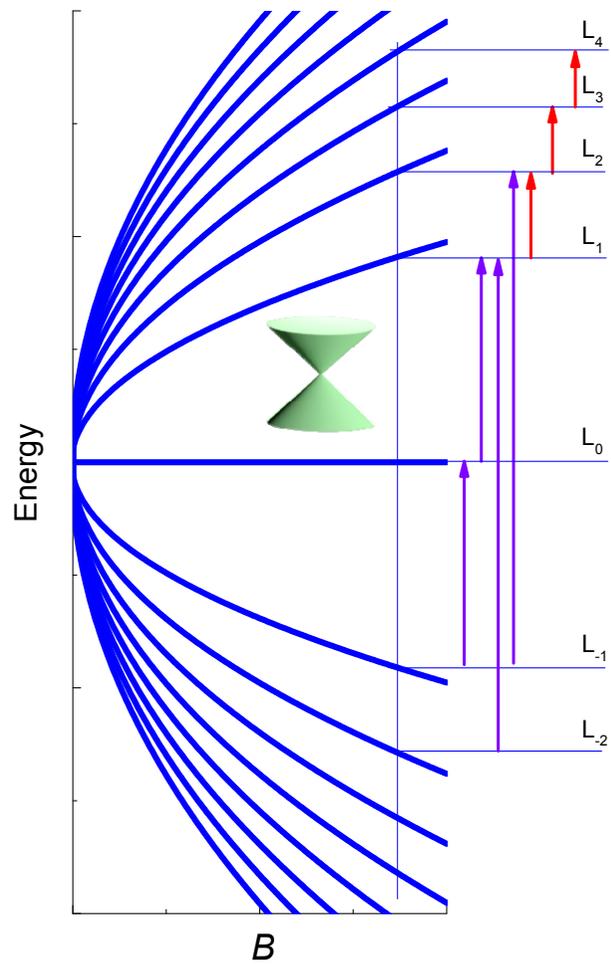
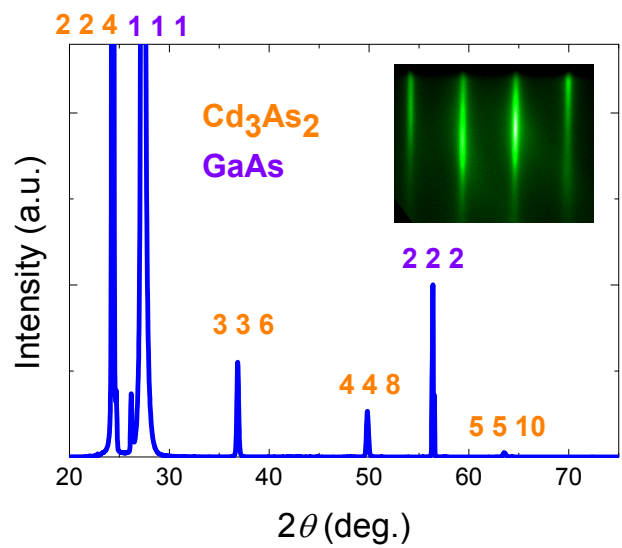
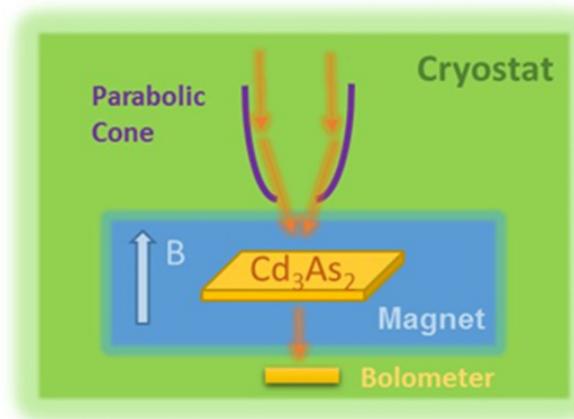
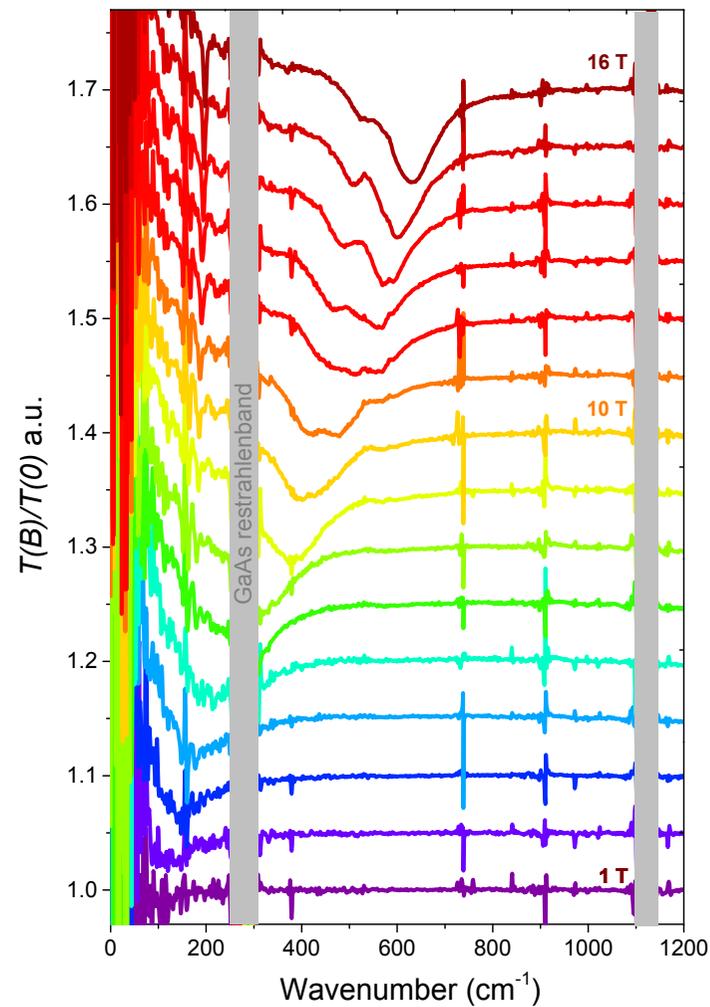

X. Yuan, et al. Figure 1

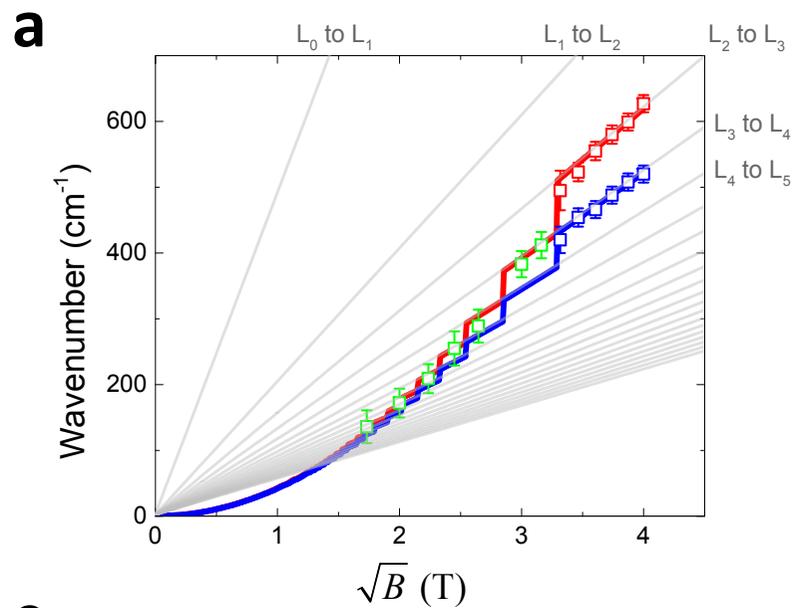
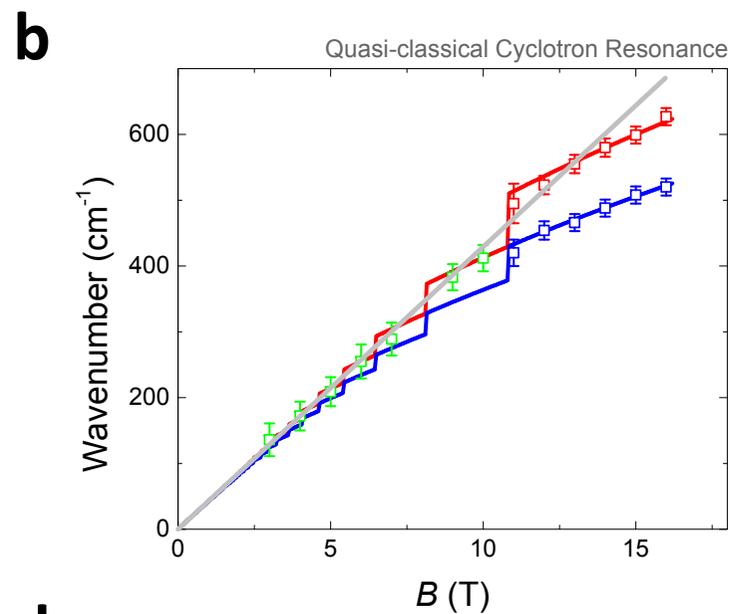
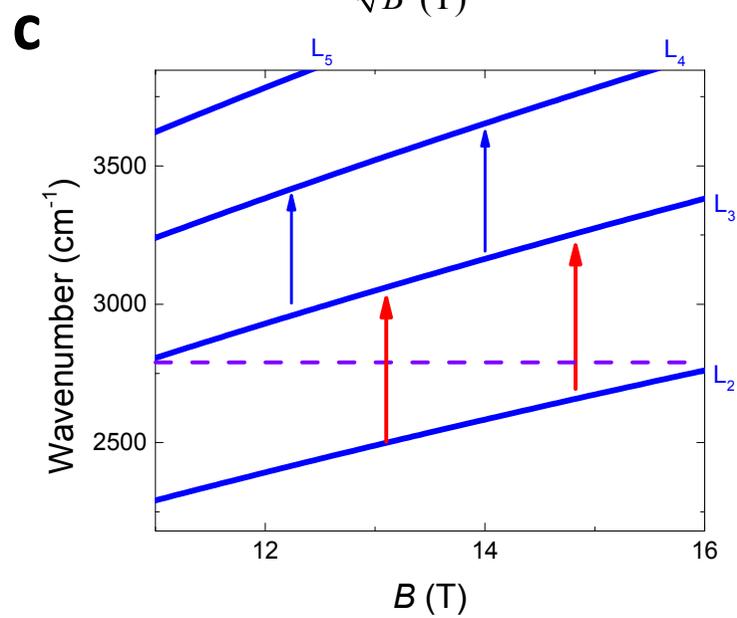
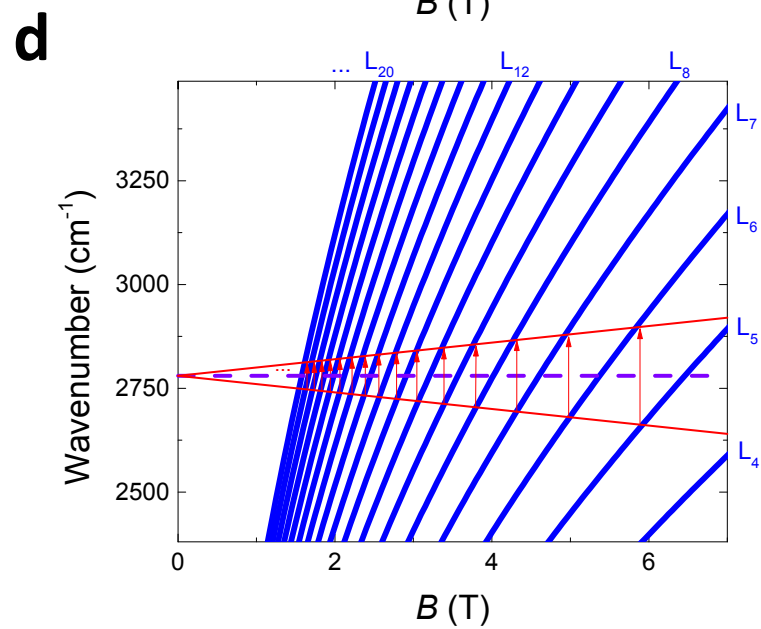

X. Yuan, et al. Figure 2

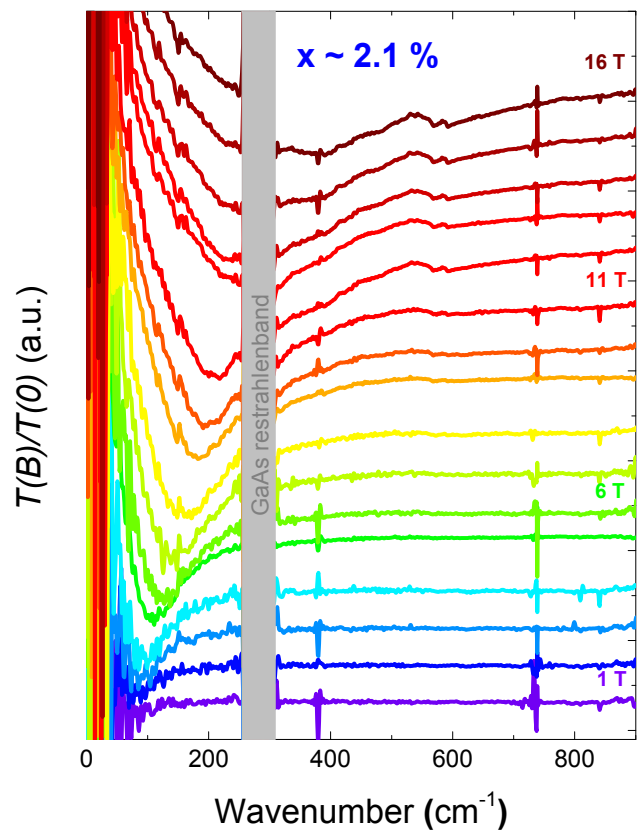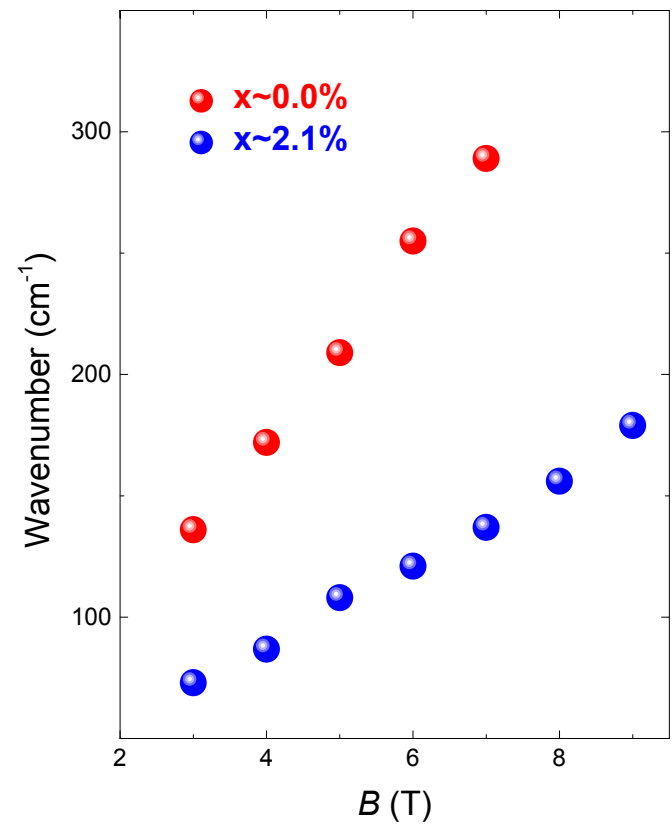

X. Yuan, et al. Figure 3

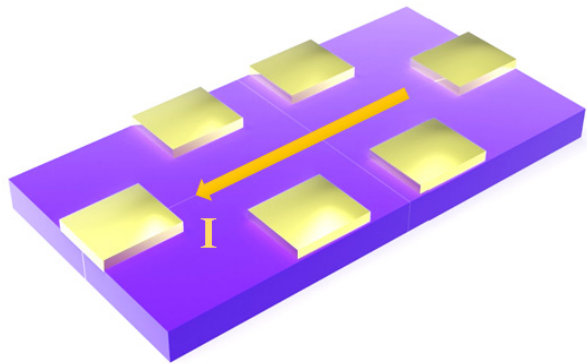
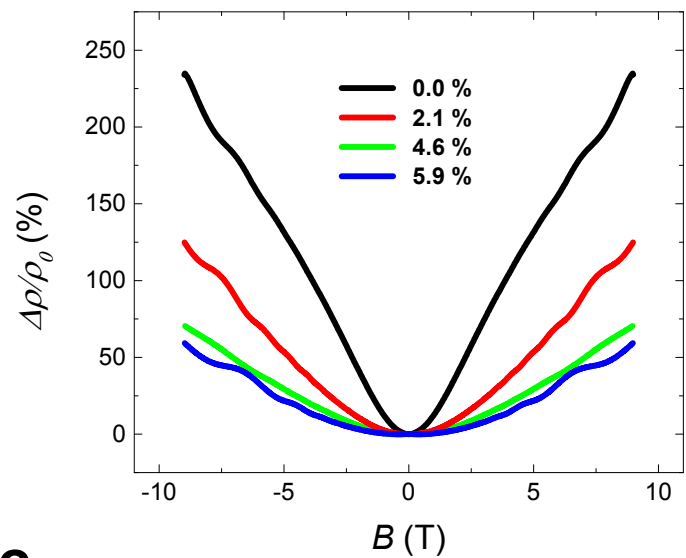
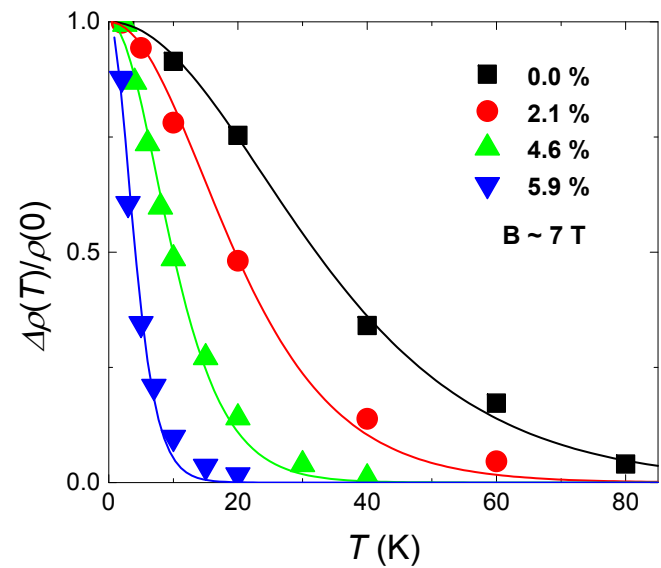
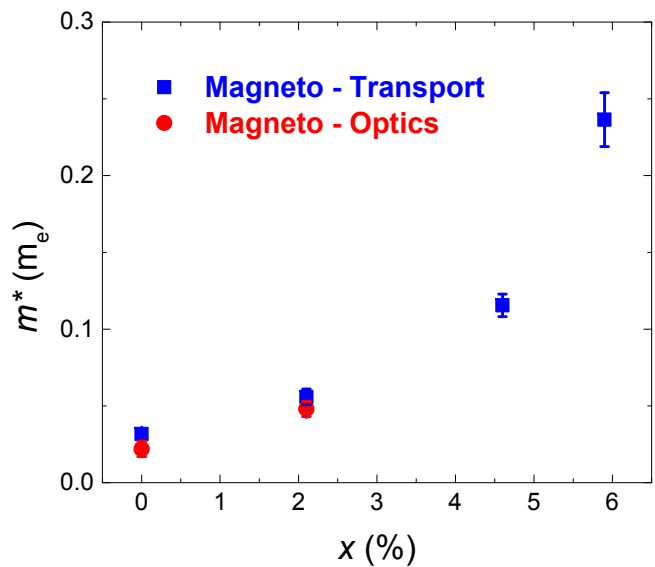
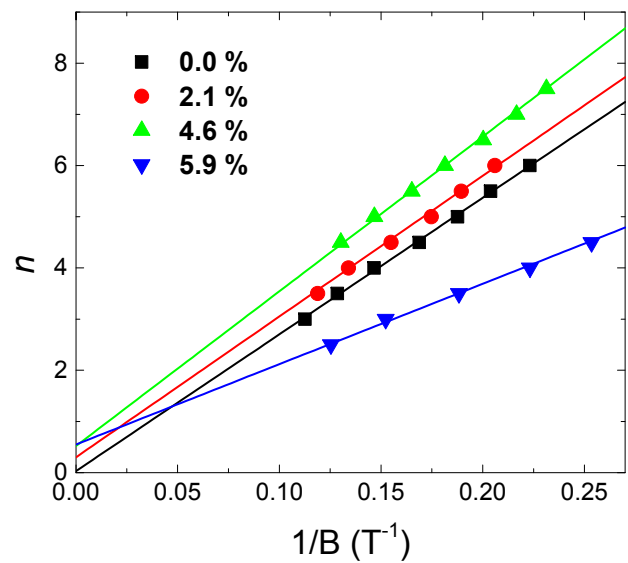
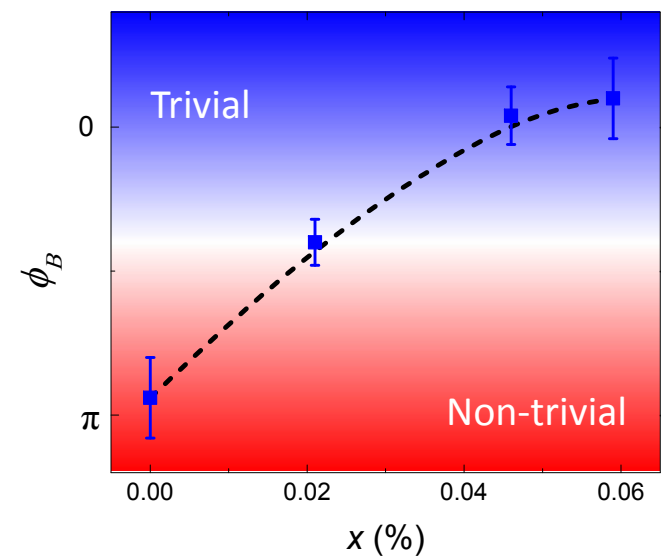

X. Yuan, et al. Figure 4

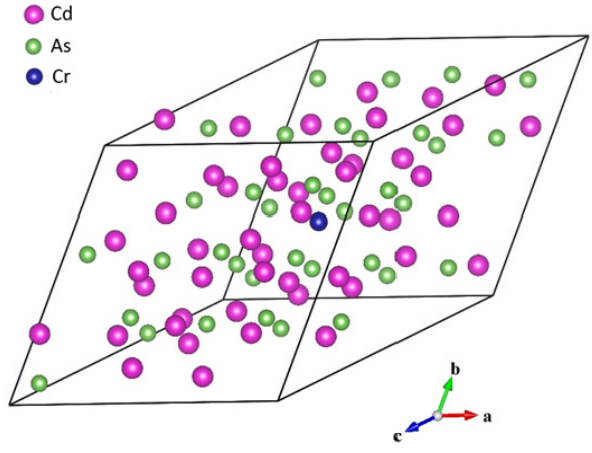 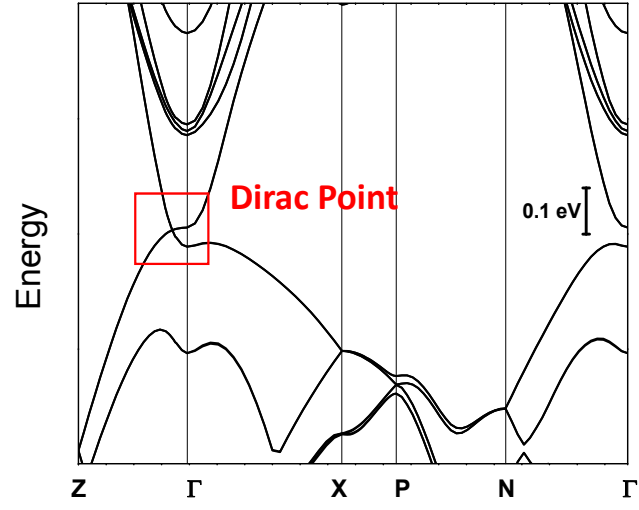 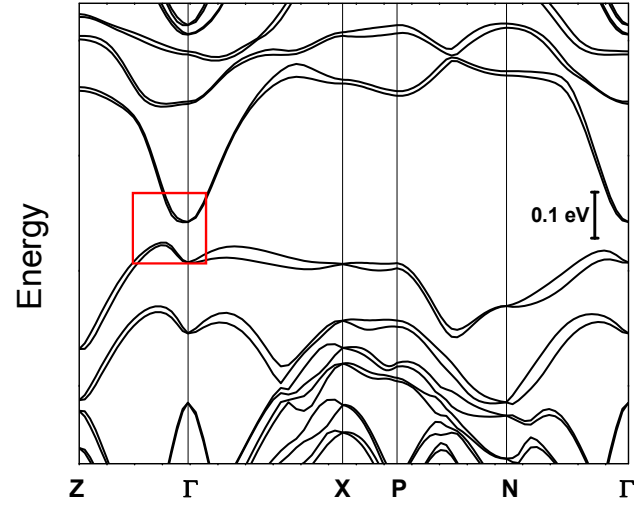

X. Yuan, et al. Figure 5